\documentclass[preprints,article,accept,moreauthors,pdftex]{Definitions/mdpi} 
\firstpage{1} 
\pubvolume{xx}
\issuenum{1}
\articlenumber{5}
\pubyear{2019}
\copyrightyear{2019}
\history{Dated: \today}

\pdfoutput=1

\usepackage{amssymb}
\usepackage{tikz}
\usetikzlibrary{positioning}
\usepackage{dsfont} 
\newcommand{\mathH}{\hat{\mathcal{H}}}

\Title{The Effect of a Positive Cosmological Constant on the Bounce of Loop Quantum Cosmology}
\Author{Mercedes Mart\'{i}n-Benito * $^{,\dagger}$ and Rita B. Neves $^{\dagger}$}

\AuthorNames{Mercedes Mart\'{i}n-Benito and Rita B. Neves}

\address[1]{%
Departamento de F\'isica Te\'orica and IPARCOS,  Universidad Complutense de Madrid,  Parque de Ciencias 1,   28040 Madrid, Spain}

\corres{Correspondence: m.martin.benito@ucm.es}

\firstnote{These authors contributed equally to this work.}

\abstract{We provide an analytical solution to the quantum dynamics of a flat Friedmann-Lema\^itre- Robertson-Walker model with a massless scalar field in the presence of a small and positive cosmological constant, in the context of Loop Quantum Cosmology. We use a perturbative treatment with respect to the model without a cosmological constant, which is exactly solvable. Our solution is approximate, but it is precisely valid at the high curvature regime where quantum gravity corrections are important. We compute explicitly the evolution of the expectation value of the volume. For semiclassical states characterized by a Gaussian spectral profile, the introduction of a positive cosmological constant displaces the bounce of the solvable model to lower volumes and to higher values of the scalar field. These displacements are state dependent, and in particular, they depend on the peak of the Gaussian profile, which measures the momentum of the scalar field. Moreover, for those semiclassical states, the bounce remains symmetric, as in the vanishing cosmological constant case. However, we show that the behavior of the volume is more intricate for generic states, leading in general to a non-symmetric bounce.}

\keyword{loop quantum cosmology, quantum bounce, cosmological constant, early Universe physics}

\begin{document}

\section{Introduction}

In the last decade, the field of quantum cosmology has experienced a surge in activity, driven in part by the idea that the physics of the early Universe might reveal genuine quantum gravity effects. Together with the increasingly more accurate observations and analyses of the Cosmic Microwave Background (CMB), this has lead to the fairly realistic hope of finding an observational window for quantum gravity. In fact, the latest analysis of the power spectrum of the CMB, although in general agreeing with the predictions from standard cosmology, reveals tensions between predictions and observations at large angular scales \cite{Ade_2015PlanckResults,Ade_2015PlanckResults2}. It is thought that these anomalies could be the result of physical processes occurring in the very early Universe, and thus may be due to genuine quantum gravity effects on the primordial cosmological perturbations.

Investigations into the consequences of quantum cosmology on the CMB have been carried out within various formalisms, including Loop Quantum Cosmology (LQC) and quantum geometrodynamics~\cite{Kiefer:2012,Brizuela:2016}. We focus our comments on LQC. LQC is the application of the techniques of Loop Quantum Gravity (LQG) to cosmological models \cite{LQCreview_Bojowald2005,LQCreview_Ashtekar2011, LQCreview_Banerjee2012, LQCreview_Agullo2016}. It has been successfully applied to homogeneous isotropic Friedmann-Lema\^itre-Robertson-Walker (FLRW) spacetimes \cite{APS_PRL,APS_extended,Bentivegna2008,Kaminski2009,Pawlowski2012}, leading to the remarkable replacement of the big-bang singularity with a quantum bounce. Inhomogeneous models, such as Gowdy models, have also been studied within this context, resorting to a hybrid quantization strategy in which loop methods are used for the zero-modes of the geometry, and Fock techniques for the fields describing the inhomogeneities \cite{Merce_2008Hybrid,Garay_2008hybrid,Merce_2010hybrid}. Recently, this approach has been successfully applied to cosmological perturbations in inflationary scenarios as well \cite{Mendez_2012hybridInflation,Mendez_2014hybridInflation,Gomar_2014hybridInflation,Gomar_2015hybridInflation}. In particular, the works of \cite{Gomar_2014hybridInflation,Gomar_2015hybridInflation} analyze the perturbations of an FLRW spacetime minimally coupled to a scalar field (an inflaton) subject to a potential. Using this hybrid approach and a type of Born-Oppenheimer ansatz for the physical states, the quantum equation that dictates the dynamics of the cosmological perturbations is derived. It turns out that this equation is in fact the Mukhanov-Sasaki (MS) equations with corrections that encode quantum gravity effects. In \cite{Gomar_2015hybridInflation} the derivation of the modified MS equations was generalized by describing the whole system with gauge invariant variables, including the zero-modes of the FLRW geometry, and without specifying the quantization adopted for these modes. These modified MS equations can then be used to compute the power spectrum of the primordial perturbations, thereby extracting predictions, in particular from LQC, which can be contrasted with observations.

So far, this analysis has only been performed after introducing semiclassical or effective approximations for the description of the FLRW geometry, by replacing expectation values on FLRW states with classical values evolved with effective dynamics of homogeneous and isotropic LQC. The~corresponding modifications to the large angular scales of the scalar power spectrum have been studied in detail (see, e.g., \cite{Agullo:2013ai, Agullo_2015PS,Agullo_2015CMB, Ashtekar:2016wpi, deBlas:2016puz, Gomar:2017yww}). Nevertheless, it would also be desirable to go beyond these effective approximations, thereby retaining the quantum nature of the background FLRW geometry.

The work in \cite{Merce_MSeq} proposed a procedure to compute the quantum corrections at the Planck regime to the MS equations  up to the maximum practical extent. Thus, a falsifiable picture of LQC is possible by means of CMB observations, as this procedure would allow for a proper discrimination between various formalisms and prescriptions within quantum cosmology, and in particular, LQC. The goal is to obtain better quantitative predictions for the CMB power spectrum, which could reveal new phenomena. This~way, it is also possible to check that additional quantum corrections are indeed negligible in scenarios where they have been neglected; e.g., when effective dynamics for the description of the FLRW geometry are taken in the place of the full quantum dynamics.

This procedure tackles the issue of the quantum evolution of each FLRW state, as in general, in the presence of a non vanishing field potential, it is determined by a time-dependent quantum Hamiltonian which is not integrable. With this method, the potential is viewed as a kind of geometric interaction. One~first takes the free geometric part of the generator of the FLRW dynamics (corresponding to a vanishing potential), and uses it to pass to an interaction picture. The evolution generated by the free geometric operator can be integrated explicitly, and even analytically by using the so-called solvable LQC (sLQC) formulation. On the other hand, the potential is regarded as a perturbation to the free dynamics. Its~dominant contributions are extracted and passed to a new interaction picture, where the corresponding evolution operator is expanded in powers of the potential, which is truncated at the desired order. If~necessary, the remaining evolution can be accounted for with a semiclassical or effective treatment.

In this study, we apply this procedure to the simplest case of a constant positive potential, namely, a positive cosmological constant, and work within linear perturbative order. We take the potential to be positive to have a toy model for inflation (even if eternal), although all our considerations apply  for a negative one as well. 
Even though in this scenario the quantum Hamiltonian is not time-dependent, the~corresponding FLRW states do not admit a simple analytically closed form, and heavy computational power is required to generate them numerically, as has been done in \cite{Pawlowski2012}. Hence, this model provides the perfect test of the procedure. Namely, it allows for the development of the required mathematical tools in a simplified scenario, and the final result can be checked against one of the exact (but more time and resource-consuming) treatment. Indeed, the dynamics of this model obtained with our treatment agrees with that of \cite{Pawlowski2012}. However, we are able to take the analysis further by comparing the evolution of the models with and without cosmological constant. We conclude that, for semiclassical states characterized by a Gaussian profile, the introduction of a positive cosmological constant pushes the bounce to lower volumes and higher values of the scalar field. Furthermore, we are able to uncover interesting behaviors of the dynamics for generic states. In fact, we find that the simple behavior of a symmetric bounce is particular to semiclassical states with Gaussian profiles, but that the dynamics of generic states are more intricate, with terms that are not symmetric. This displays clearly one advantage of this perturbative method: by keeping only the leading order contributions of the potential to the dynamics, we simplify the computations and do not have to particularize the investigation to semiclassical states or to rely on numerical treatments. This allows us to retain more information and uncover effects that might otherwise be ignored.

As our work is seen as a stepping stone to the application of this perturbative method to other models (namely, inflationary ones), we keep the calculations general as much as possible, only particularizing for a constant potential when it is necessary to move forward in the calculations.

The paper is organized as follows. In Section \ref{sec:sLQC} we review the usual procedures of solvable LQC for the quantization of an FLRW model minimally coupled to a massless scalar field, as it is fundamental for the method that we apply. In Section \ref{sec:generic potential}, we review the procedure proposed in \cite{Merce_MSeq}, and summarize the calculations that lead to the expectation value of the volume operator. In Section \ref{sec:constant potential} we apply the treatment for the case of a constant potential, obtaining the expectation value of the volume, and compare it with classical trajectories. In Section \ref{sec:bounce with potential} we compare the scenarios with and without cosmological constant, and~investigate further the effect of introducing a cosmological constant in the bounce scenario of the solvable model. Section \ref{sec:approx} is devoted to discussing the regime of the validity of our approximation. Lastly, in Section \ref{sec:discussion} we conclude with a discussion of our work.

We choose units $c=\hbar=1$.

\section{Homogeneous and Isotropic LQC}\label{sec:sLQC}

In this section, we review the usual procedures of LQC for the quantization of an FLRW model minimally coupled to a scalar field. At first we allow for a non vanishing field potential $W(\phi)$; however, we  review the solvable formulation of the case of vanishing potential, as it  essential for the remaining work. For more details on the procedures of this section see  \cite{APS_extended, ACS}.

Following the techniques of LQG, in LQC the gravitational sector of the system is described with a $SU(2)$ connection, namely, the Ashtekar-Barbero connection, and its canonically conjugate densitized triad. In LQG, there is no operator directly representing the connection, and holonomies of the connection along with fluxes of the densitized triad are used as fundamental variables. Given homogeneity and isotropy, in~a flat FLRW model the connection can be written as proportional to a spatially constant variable $c$ and the densitized triad to a spatially constant variable $p$. One usually restricts the study to a finite spatial cubical cell $\mathcal V$, whose volume is $V_o$ when measured with a fiducial Euclidean metric ${}^o q$. The relation of $p$ with the scale factor is given by $a(t)=\sqrt{|p(t)|}V_o^{-1/3}$, and the sign of $p$ depends on the relative orientation between the physical triad and the fiducial triad compatible with ${}^o q$. This way, the variables $c$ and $p$ form a canonical pair with Poisson brackets $\{c,p\}=8\pi G\gamma/3$, where $\gamma$ is the so-called Immirzi parameter and $G$ is the Newton constant. Following the "improved dynamics" prescription \cite{APS_extended}, the holonomies of the connection are taken along straight lines in the fundamental representation of $SU(2)$ with a length such that the physical area enclosed by them is $\Delta$, the gap of the area operator in LQG, whereas the fluxes are simply proportional to $p$.

To simplify calculations, one performs the following change of variables \cite{APS_extended}:
\begin{align}
	v &= \text{sign}(p) \frac{|p|^{3/2}}{2 \pi G \gamma \sqrt{\Delta}},\quad
	b = \sqrt{\frac{\Delta}{|p|}} c,
\end{align}
with $\lbrace b,v \rbrace=2$. In this new canonical set, the holonomies simply produce a constant shift in $v$, the new geometric variable that replaces $p$. The volume of the cell $\mathcal V$ is then given by $V = 2\pi G \gamma \sqrt{\Delta}|v|$.

The matter content is represented by a scalar field $\phi$ and its canonically conjugate momentum $\pi_{\phi}$, with $\{\phi,\pi_{\phi}\}=1$. Classically, this system satisfies the constraint:
\begin{equation}\label{eq:classicalconstraint}
	C = \pi_{\phi}^2 - \frac{3}{4 \pi G \gamma^2} \Omega^2 + 8\pi^2 G^2\Delta \gamma^2 v^2 W(\phi) = 0,
\end{equation}
where $\Omega = 2 \pi G \gamma b v$.

Remarkably, in LQC, in the representation where the operator for $v$ (and also the one for $\phi$) acts by multiplication, the inner product is discrete in the volume variable. This is the foundation for the outstanding features of LQC, which set it apart from other quantization procedures. Furthermore, in~this representation, the physical states decouple in superlection sectors, spanned by volume eigenstates which differ by multiples of four units between them. There is an especially useful representation named sLQC \cite{ACS}, in which the free model (with vanishing potential) can be straightforwardly solved analytically. Namely, taking the sector spanned by the states supported on the volume variable $v$ equal to a multiple of four ($v=4n$, where $n$ is an integer), a discrete Fourier transform is performed from $v$ to $b$, followed by a rescaling of $\Gamma$, the wave functions of the $(v,\phi)$ representation, by $\chi=\Gamma/(\pi v)$ and the change of variable:
\begin{equation}
	x=\frac{1}{\sqrt{12\pi G}}\ln\left[ \tan\left( \frac{b}{2} \right) \right],
\end{equation}
so that $b=2\tan^{-1} \left( e^{\sqrt{12\pi G}x} \right)$. This way, representing the functions of the connection in terms of holonomy elements $e^{ib/2}$, we obtain a counterpart of $\Omega^2$ which is equivalent to replacing $b$ with $\sin b$. Thus, in the particular case of vanishing potential $W(\phi)=0$, the quantum counterpart of the constraint \eqref{eq:classicalconstraint} reads
\begin{equation}\label{eq:constraint_x}     (\hat{\pi}_{\phi}^2-\hat{\pi}_x^2)\chi(x,\phi)=0,
\end{equation}
where $\hat{\pi}_x$ and $\hat{\pi}_{\phi}$ are the two momentum operators, which act by derivative, namely, $\hat{\pi}_x=-i\partial_x$ and $\hat{\pi}_\phi=-i\partial_\phi$. We also have $\hat{\pi}_x = \sqrt{3/(4\pi G\gamma^2)}\hat{\Omega}$. Now, we regard $\phi$ as (internal) time, and thus the constraint is a Klein-Gordon equation in 1+1 dimensions. To remove double counting of solutions due to time reversal invariance, we take $\hat{\pi}_{\phi}$ to be positive. While in the kinematical Hilbert space the operators $\hat\phi$ and $\hat\pi_\phi$ provide a pair of canonically conjugated essentially self-adjoint operators, note that at the physical level, $\phi$ is the time variable (and it has no associated operator defined in the physical Hilbert space) and $\hat\pi_\phi=\pm|\hat\pi_x|$ is the Hamiltonian generating the evolution of positive and negative frequency solutions respectively. The positive frequency solutions of the constraint \eqref{eq:constraint_x} are then functions of the form $\chi(x_{\pm}) = \chi(\phi \pm x)$, which correspond to left and right moving modes, respectively. Here $\chi$ is any function whose Fourier transform is supported on the positive real line. The operator $\hat{\pi}_{x}=-i\partial_x$ is positive on left-moving modes and negative on right-moving modes. Moreover,
invariance under reversal of the triads imposes the condition that  physical states verify $\chi(x,\phi)=-\chi(-x,\phi)$, which leads to physical states of the form:
\begin{equation}\label{eq:states_sLQC}
	\chi(x,\phi) = \frac{1}{\sqrt{2}}\left[\chi(x_+)-\chi(x_-)\right].
\end{equation}

This way, left and right moving modes are not independent (as would be the case had we performed a standard quantization instead of the polymeric one of LQC), and we can write physical results using only, e.g., left-moving modes. In particular, the (time-independent) inner product on physical states can be written as
\begin{equation}\label{eq:innerprod_sLQC}
    ( \chi_1,\chi_2 )=2 i \int_{-\infty}^\infty dx\ [\partial_x{\chi_1}^*(x_+)] \chi_2(x_+),
\end{equation}
where $*$ denotes complex conjugation.

The dynamics can be integrated straightforwardly: in Schr\"odinger picture, the evolution of positive frequency states from an initial time $\phi_0$ is given by 
\begin{equation}
    \chi(x,\phi)=e^{i|\hat{\pi}_x|\Delta\phi}\chi_0(x),
\end{equation}
$\chi_0$ being the initial FLRW state at $\phi_0$, and $\Delta\phi=\phi-\phi_0$. In what follows we choose 
\begin{equation}\label{eq:chi0}
    \chi_0(x)=\frac{1}{\sqrt{2}}\left[F(x)-F(-x)\right],
\end{equation}
so that 
$\chi(x,\phi) = \left[F(\Delta\phi+x)-F(\Delta\phi-x)\right]/{\sqrt{2}}$.
We recall that $F$ is a function with Fourier transform supported on the positive real line as we are restricting to the positive frequency sector.

Moreover, employing Heisenberg picture, $\hat{\pi}_x$ and $\hat{\pi}_{\phi}$ are Dirac observables, preserved by the dynamics, whereas $\hat{x}(\phi)$ is found to be
\begin{equation}\label{eq:xevolution}
 \hat{x}(\phi)= \hat{x}_0 - \Delta\phi\, \text{sign}(\hat{\pi}_x).
\end{equation}

Here $\hat{x}_0$ represents $\hat{x}$ in the section of evolution where $\phi=\phi_0$.
Furthermore, the operator that represents the volume of the cell $\mathcal V$ is given (in Schr\"odinger picture) by \cite{ACS}:
\begin{equation}
   \hat{V}=\frac{2\pi G\gamma \sqrt{\Delta}}{\sqrt{3\pi G}} \cosh(\sqrt{12\pi G}\hat{x}) |\hat{\pi}_x|.
\end{equation}

The Heisenberg picture counterpart, $\hat{V}(\phi)$, is obtained simply by replacing $\hat{x}$ with $\hat{x}(\phi)$. Then,  as a function of the internal time $\phi$, the expectation value of the volume operator on all physical states turns out to be
\begin{equation}\label{eq:expVsLQC}
	\langle \hat{V}(\phi)\rangle_{\chi_0}= 2 \pi G \gamma \sqrt{\Delta}\left(v_+e^{\sqrt{12\pi G}\ \Delta\phi}+v_-e^{-\sqrt{12\pi G}\ \Delta\phi}\right),
\end{equation}
 where $\langle \hat{V}(\phi)\rangle_{\chi_0}\equiv({\chi_0}, \hat{V}(\phi){\chi_0} )=({\chi}, \hat{V}{\chi} )$ and $v_\pm$ are state-dependent constants defined, in terms of the function $F(x)$, as
\begin{equation}\label{eq:v+-sLQC}
	v_\pm \equiv \frac{1}{\sqrt{3\pi G}}\int_{-\infty}^{+\infty} dx \left| \frac{d{F}(x)}{dx} \right|^2 e^{\mp\sqrt{12\pi G}\ x}.
\end{equation}

\section{Corrections from a Field Potential in LQC}\label{sec:generic potential}

In this section, we summarize the procedure proposed in \cite{Merce_MSeq} to obtain further corrections to the quantum dynamics of the model with a generic field potential $W(\phi)$. Taking $W=0$, we would obtain the trajectories \eqref{eq:xevolution} and the analytical expression for the states \eqref{eq:states_sLQC} along with the physical inner product~\eqref{eq:innerprod_sLQC}. With this procedure, we take the generator of the FLRW dynamics, extract first its free geometric part ($W=0$) and use it to pass to an interaction picture. Then, a perturbative treatment allows us to obtain the dominant contributions of the potential to the dynamics.

We keep contributions to first order in $W$, and show explicitly the calculations that lead to the expectation value of the volume operator, so as to simplify future applications of this procedure to other~models.

\subsection{Procedure}

Let us now include a field potential in this description. The quantum operator that represents the constraint \eqref{eq:classicalconstraint} is
\begin{align}
    \hat{C} &= \hat{\pi}_{\phi}^2-\mathH_0^{(2)},\qquad
    \mathH_0^{(2)} = \hat{\pi}_x^2 - 2 W({\hat\phi})\hat{V}^2.
\end{align}

Now, at the physical level, the quantum Hamiltonian is $\mathH_0 = \hat{\pi}_{\phi}$, and generates evolution in the time parameter $\phi$. Considering only positive frequency solutions, $\mathH_0$ is given by the positive square root of $\mathH_0^{(2)}$, and is seen as a perturbation of the free system Hamiltonian $\mathH_0^{(F)} = |\hat{\pi}_x|$. In this spirit, we write the states as $\chi(x,\phi) = \hat{U}(x,\phi)\chi_0(x)$, where $\chi_0$ is the initial FLRW state at $\phi=\phi_0$ of the free system previously introduced, and $\hat{U}$ is the evolution operator
\begin{equation}
    \hat{U}(x,\phi)=\mathcal{P}\left[\exp\left(i \int_{\phi_0}^{\phi} d\tilde{\phi} \mathH_0(x,\tilde{\phi})\right)\right],
\end{equation}
where $\mathcal{P}$ denotes time ordering (with respect to $\phi$). Thus, expectation values are taken on the state $\chi$ of the FLRW geometry, with the inner product of sLQC \eqref{eq:innerprod_sLQC}. Now, we pass to an interaction picture, where any operator $\hat{A}$ of the original Schr\"odinger-like picture has its counterpart:
\begin{equation}
    \hat{A}_I(\phi) = e^{-i\mathH_0^{(F)}\Delta\phi} \hat{A} e^{i\mathH_0^{(F)}\Delta\phi}.
\end{equation}
Writing $\mathH_{1} = \mathH_0 - \mathH_0^{(F)}$, the states in this picture can then be described by
\begin{align}
    \chi_I(x,\phi) &= \hat{U}_I(x,\phi)\chi_0(x),\\
    \hat{U}_I(x,\phi) &= \mathcal{P}\left[ \exp\left( i \int_{\phi_0}^{\phi} d\tilde{\phi} \mathH_{1I}(x,\tilde{\phi})\right) \right].
\end{align}

This leads to the following form for the expectation value of an operator:
\begin{equation}
   \langle \hat{A} \rangle_{\chi}= \langle \hat{A}(\phi) \rangle_{\chi_0} = \langle \hat{U}_I^{\dagger}(\phi) \hat{A}_I(\phi) \hat{U}_I(\phi) \rangle_{\chi_0},
\end{equation}
where the dagger denotes an adjoint. Note that, given the dynamics of the free case, the forms of operators of the full system in the interaction picture are simply found by replacing their dependence on $\hat{x}$ by the same dependence on the evolved operator $\hat{x}(\phi)$ defined in equation \eqref{eq:xevolution}. Therefore, the obstacle is reduced to the computation of $\hat{U}_I$.

Assuming we can regard the potential as a perturbation of the free system, we extract from $\hat{U}_I$ the dominant contributions of the potential. Up to first order, and given a suitable factor ordering, it is shown in \cite{Merce_MSeq} that $\mathH_1$ can be represented approximately as
\begin{equation}\label{eq:H2}
    \mathH_1 \simeq \mathH_2 = -W(\phi) \hat{B},
\end{equation}
where the auxiliary operator $\hat{B}$ is defined as
\begin{equation}\label{eq:B}
    \hat{B} = \frac{4\pi G\Delta \gamma^2}{3}\cosh^2\left(\sqrt{12\pi G}\hat{x}\right)|\hat{\pi}_x|.
\end{equation}

Generally, we can write $\mathH_{1I} = \mathH_{2I} + \mathH_{3I}$, where $\mathH_{2I}$ is the operator \eqref{eq:H2} in the interaction picture and $\mathH_{3I}$ is the remaining part of $\mathH_{1I}$, at least of second order in the potential. The dynamics generated by $\mathH_{2I}$ can be obtained by passing to a new interaction picture $J$, for which we introduce:
\begin{equation}
    \hat{U}_{2I}(x,\phi) = \mathcal{P}\left[ \exp\left( i \int_{\phi_0}^{\phi} d\tilde{\phi} \mathH_{2I}(x,\tilde{\phi})\right) \right].
\end{equation}
To any operator $\hat{A}_I$ in the original interaction picture corresponds the operator $\hat{A}_J$ of the new interaction picture:
\begin{equation}
    \hat{A}_J(\phi) = \hat{U}_{2I}^{\dagger}(\phi) \hat{A}_I(\phi) \hat{U}_{2I}(\phi). 
\end{equation}

Finally, this leads to the expectation value of any operator $\hat{A}$ of the Schr\"odinger-like picture on states of the FLRW geometry to be given by
\begin{equation}
    \langle \hat{A}(\phi) \rangle_{\chi_0} = \langle \hat{U}_J^{\dagger}(\phi) \hat{A}_J(\phi) \hat{U}_J(\phi) \rangle_{\chi_0},
\end{equation}
where
\begin{equation}
    \hat{U}_J = \mathcal{P}\left[ \exp\left( i \int_{\phi_0}^{\phi} d\tilde{\phi} \mathH_{3J}(x,\tilde{\phi})\right) \right].
\end{equation}

Up to this point the treatment has been exact. However, we now need to introduce approximations. Firstly we consider situations for which the evolution generated by $\mathH_3$ is negligible and can be ignored, and thus $\hat{U}_J \simeq \mathds{1}$ and
\begin{equation}
    \langle \hat{A}(\phi) \rangle_{\chi_0} \simeq \langle \hat{A}_J(\phi) \rangle_{\chi_0}.
\end{equation}

Secondly, we expand the path ordered integral in $\hat{U}_{2I}$ in powers of the potential and truncate at first~order:
\begin{equation}
    \hat{U}_{2I} = \mathds{1} + i \int_{\phi_0}^{\phi} d\tilde{\phi} \mathH_{2I}(x,\tilde{\phi}) + \mathcal{O}(W^2).
\end{equation}

Keeping only linear contributions of the potential yields
\begin{equation}\label{eq:Aj}
    \hat{A}_J(\phi) \simeq \hat{A}_I(\phi) - i\left[\hat{A}_I\phi),\int_{\phi_0}^{\phi} d\tilde{\phi}\,W(\tilde\phi)\hat{B}_I(\tilde{\phi})\right].
\end{equation}

This way, given the inner product of sLQC \eqref{eq:innerprod_sLQC} and an initial state $\chi_0$, we are ready to compute the expectation value of any operator, to first order in the potential.

\subsection{Expectation value of the volume operator}\label{sec:expV_constantW}

Generally, we are interested in tracking the expectation value of the volume operator in the model under consideration:
\begin{equation}\label{eq:Vexp1}
    \langle \hat{V}(\phi) \rangle_{\chi_0}\simeq \langle \hat{V}_J(\phi) \rangle_{\chi_0} =2i\int_{-\infty}^\infty dx \frac{dF^*(x)}{dx}\hat{V}_J(\phi)F(x).
\end{equation}

Let us show this calculation with generic potential explicitly, so that this task is simplified when studying particular forms of the potential in the future.
To simplify notation, we define
\begin{equation}\label{eq:I1}
    I_1(x,\phi) \equiv \int_{\phi_0}^{\phi} d\tilde{\phi}\,W(\tilde\phi)\cosh^2\left[\sqrt{12\pi G}(x-\Delta\tilde\phi)\right],
\end{equation}
with $\Delta\tilde\phi=\tilde\phi-\phi_0$.
This integral will be particular to each form of the potential $W(\phi)$. This way, $I_1$ and its derivatives are of first order in $W$. Applying \eqref{eq:Aj} for $\hat{V}_J(\phi)$, with $\hat{V}_I(\phi)$ the volume operator of the free model,
and recalling that $\hat{\pi}_x$ is positive for left-moving modes, we obtain
\begin{equation}\label{eq:VexpW}
    \langle \hat{V}(\phi) \rangle_{\chi_0} \simeq \frac{4\pi G\gamma\sqrt{\Delta}}{\sqrt{3\pi G}} \int_{-\infty}^{+\infty} dx \left| \frac{dF(x)}{dx} \right|^2 E(x,\phi),
\end{equation}
to first order in $W$, where
\begin{align}\label{eq:E}
\begin{split}
    E(x,\phi) &= \cosh\left[\sqrt{12\pi G}(x-\Delta\phi)\right]+\frac{4\pi G\gamma^2\Delta}{3}\bigg\{2\sqrt{3\pi G} I_1(x,\phi)\sinh\left[\sqrt{12\pi G}(x-\Delta\phi)\right] \\
    &- \cosh\left[\sqrt{12\pi G}(x-\Delta\phi)\right] \partial_x I_1(x,\phi)\bigg\}.
\end{split}
\end{align}

For specific forms of the potential, it  useful to identify in $\langle \hat{V}(\phi) \rangle_{\chi_0}$ the constants
\begin{equation}\label{eq:vn}
	v_n \equiv \frac{1}{\sqrt{3\pi G}}\int_{-\infty}^{+\infty} dx \left| \frac{dF(x)}{dx} \right|^2 e^{-n\sqrt{12\pi G}\ x},
\end{equation}
in order to express $\langle \hat{V}(\phi) \rangle_{\chi_0}$ in a manner analogous to that of \eqref{eq:expVsLQC}. Note that $v_\pm$ defined in \eqref{eq:v+-sLQC} corresponds to taking $v_{\pm 1}$, and so we verify that for $W=0$ ($I_1=0$) we recover the case of sLQC, as we should.

In what follows, to simplify notation, $\langle \hat V \rangle$ means $\langle \hat{V} \rangle_{\chi}=\langle \hat{V}(\phi) \rangle_{\chi_0}$, and likewise for other operators.

\section{Constant Potential}\label{sec:constant potential}

This procedure has been proposed in \cite{Merce_MSeq}, but so far, it has not been applied to a specific form of the potential. In this section we do so for the simplest form of the potential: a constant positive one. Even though this is a simple scenario, it is already a relevant model for cosmology, as it is equivalent to considering a massless scalar field in the presence of a positive cosmological constant $\Lambda$, having $W = \Lambda / 8\pi G$. It is also relevant as a test model for the procedure, before we implement it with other more complicated forms of $W$. 

A constant potential provides a time-independent Hamiltonian. This way, it allows for a simplification of the calculations, while still enabling the development of techniques that will be useful for the analysis of other potentials. 
Additionally, the loop quantization of this model has been rigorously carried out in \cite{Pawlowski2012}, and the quantum dynamics of a family of semiclassical states determined by Gaussian profiles has been obtained by generating numerically the eigenfunctions of the Hamiltonian. This provides a check for the validity of our method. 

Classically, the model presents two types of solutions: universes expanding from a big bang to infinite volume (reached at infinite proper time), and universes contracting from infinite volume to a big crunch (reached at infinite proper time). In internal time $\phi$, the expanding solutions reach infinite volume at a finite value of $\phi$, and one can analytically continue the evolution, matching it with that of the contracting solution.
Moreover, as it happens for vanishing cosmological constant, the loop quantization drastically modifies the dynamics at the Planck regime, replacing the big bang or big crunch singularities with a quantum bounce. This leads to an essentially periodic (in time $\phi$) universe \cite{Pawlowski2012}.

In this work we are interested only on the quantum modifications to the classical dynamics in the Planck regime; namely, in the quantum dynamics around the bounce. Indeed, our approximation method is only valid  when the matter energy density dominates over the energy density associated to the cosmological constant, $\Lambda/(8\pi G)$, so that the latter can be regarded as a perturbation. In consequence, we consider values of $W$ to be far smaller than the Planckian density (as discussed later, we particularly consider values such that the energy density of the cosmological constant at the bounce is  that of the scalar field times $\mathcal{O}(10^{-4})$). The main novelty of this work is a comparison between the Planck regimen dynamics of the model with $\Lambda>0$ and that of the free model ($\Lambda=0$). Namely, we analyze explicitly how the introduction of a positive (small) cosmological constant modifies the dynamics of the model without cosmological constant.

With a constant potential, it is straightforward to compute $I_1(x,\phi)$ as defined in \eqref{eq:I1}. This is followed by a manipulation of $E(x,\phi)$ of \eqref{eq:E} in order to identify the $v_n$ functions that appear in $\langle \hat{V} \rangle$ of \eqref{eq:VexpW}. These steps can be found in Appendix \ref{sec:appendix_V}. That way, we find the expectation value of the volume as
\begin{align}\label{eq:expv_W}
\begin{split}
  \langle \hat{V} \rangle \simeq 2\pi G\gamma\sqrt{\Delta}\bigg[ &\left(v_+ +\tilde{W} w_+(\Delta\phi)\right) e^{\sqrt{12\pi G}\Delta\phi}+ \left(v_- +\tilde{W} w_-(\Delta\phi)\right) e^{-\sqrt{12\pi G}\Delta\phi}\\
  &+ \tilde{W}\left(v_{+3}e^{3\sqrt{12\pi G}\Delta\phi}+ v_{-3}e^{-3\sqrt{12\pi G}\Delta\phi}\right)\bigg],
\end{split}
\end{align}
to first order in the potential, where, to simplify notation we have defined:
\begin{align}
    \tilde{W} &= W\frac{\pi G\gamma^2\Delta}{6},\label{eq:Wtilde}\\
    w_\pm(\Delta\phi) &= \left(3\mp 8\sqrt{3\pi G}\Delta\phi\right)v_\pm - v_{\pm3}-3v_\mp,
\end{align}
recalling that the constants $v_n$ are state dependent, as defined in \eqref{eq:vn}. 

So far, we have made no restriction to a particular form of the physical state $\chi$ on which we compute the expectation value. However, we are interested in tracking the expectation value of the volume on semiclassical states, by which we mean states that are highly peaked (i.e., with small relative dispersion) on the classical trajectories in the low curvature regime, where quantum effects of the geometry should be negligible. These are determined by fixing adequately the functions $F$ defined in \eqref{eq:chi0}, which in turn define the $v_n$ constants \eqref{eq:vn}. As previously mentioned, $F$ is a function with Fourier transform $\tilde{F}(k)$ supported on the positive real line:
\begin{equation}
    F(x) = \frac{1}{2\pi}\int_0^{+\infty}dk\,\tilde{F}(k)e^{i\sqrt{12\pi G}x}.
\end{equation}

For similarity with previous LQC literature, and in particular \cite{APS_extended,Pawlowski2012}, we chose semiclassical states determined by spectral Gaussian profiles (in the representation where the volume is diagonal) centered at $k_o \ll -1$ with width $\sigma \ll |k_o|$:
\begin{equation}\label{eq:gaussian profile}
	\tilde{\psi}(k) = \frac{1}{\sqrt{\sigma\sqrt{\pi}}} e^{-\frac{(k-k_o)^2}{2\sigma^2}}.
\end{equation}
These states are chosen such that the expectation value of $\hat\pi_\phi$ and its relative dispersion on them verify
\begin{equation}\label{eq:pi_phi and disp}
    \langle \hat\pi_\phi \rangle=-\sqrt{12 \pi G} k_o,\quad \frac{\langle \Delta\hat\pi_\phi \rangle}{\langle \hat\pi_\phi \rangle}=\frac{\sigma}{\sqrt{2} |k_o|}.
\end{equation}

As investigated in \cite{paper1_domainV}, the relation between the profiles	$\tilde{\psi}(k)$ of the $v$-representation and the profiles $\tilde{F}(k)$ of the solvable representation is given by
\begin{equation}
	\tilde{F}(k) = \frac{1}{\sqrt{k\pi}}\ \tilde{\psi}(-k) \cos\left(\frac{1-2ik}{4}\pi \right)\ \Gamma\left(\frac{1}{2}-ik\right),
\end{equation}
and for the states \eqref{eq:gaussian profile} we obtain \cite{paper1_domainV}
\begin{equation}\label{eq:vn_ko}
    v_n = (-k_o)^{1-n}e^{\frac{n^2}{4\sigma^2}}.
\end{equation}

Thus, for given values of $k_o$ and $\sigma$, we can compute $\langle \hat{V} \rangle$ and track its evolution along $\phi$.

\begin{figure}[H]
\centering
    \begin{tikzpicture}
    \tikzstyle{every node}=[font=\normalsize]
    \node (img)  {\includegraphics[width=0.5\textwidth]{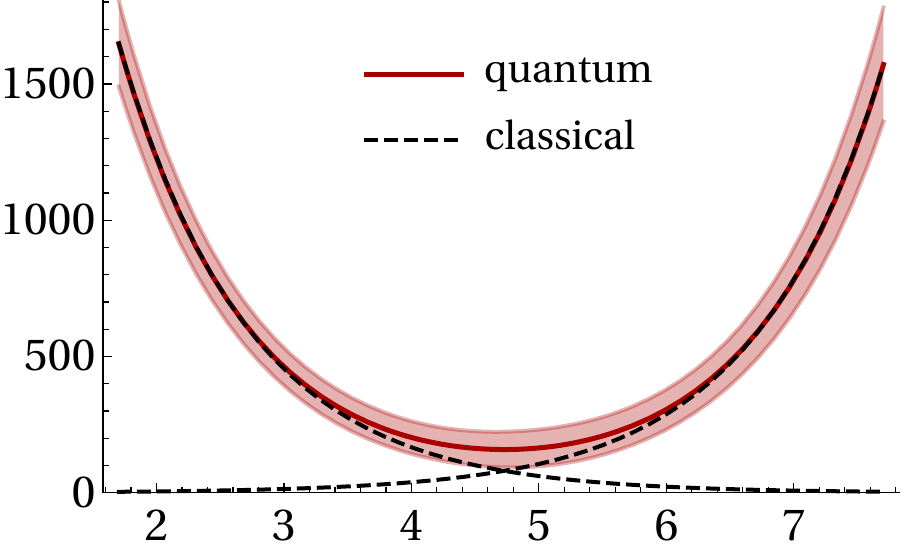}};
    \node[below=of img, node distance=0cm, yshift=1cm,] {$\sqrt{12\pi G}\Delta\phi$};
    \node[left=of img, node distance=0cm, anchor=center,rotate=90,yshift=-0.7cm,] {$V/2\pi G\gamma\sqrt{\Delta}$};
 \end{tikzpicture}
    \caption{Expectation value of $\hat{v}$ on a semiclassical state with the profile \eqref{eq:gaussian profile}, along with its dispersion (see~Appendix \ref{sec:appendix_V}), compared with the corresponding classical trajectories. The following values were used for the parameters: $k_o=-100$, $\sigma=10$ ($\langle\Delta\hat{\pi}_{\phi}\rangle/\langle\hat{\pi}_{\phi}\rangle \simeq 0.07$) and $\tilde{W} = 10^{-5}$.}
    \label{fig:dynamics}
\end{figure}

Figure \ref{fig:dynamics} shows the expectation value of $\hat{v}=\hat{V}/2\pi G\gamma\sqrt{\Delta}$ on a semiclassical state determined by \eqref{eq:gaussian profile}, and its dispersion, as  functions of time $\phi$, along with the corresponding classical trajectories. As it should be, this analysis agrees with the results of \cite{Pawlowski2012}.

The classical trajectories are obtained from the classical constraint \eqref{eq:classicalconstraint}, by finding $\dot{v}_c(\tau)=\lbrace v,C \rbrace$, $\dot{\phi}(\tau)=\lbrace \phi,C \rbrace$, where the dot denotes derivative with respect to harmonic time $\tau$, and thus
\begin{equation}
        v'_c(\phi)= \dot{v}/\dot{\phi}= \pm \sqrt{12\pi G}\left(v_c^2 +\frac{\pi G \gamma^2 \Delta\,\Lambda}{\pi_{\phi}^2}v_c^4 \right)^{1/2},
\end{equation}
where prime denotes derivative with respect to $\phi$. These differential equations can be solved numerically, resulting in the two classical trajectories: an expanding and a contracting solution.

In the low curvature regime, both the classical and quantum trajectories agree. However, close to the Planck regime, the quantum trajectory deviates from the classical ones. As classical trajectories tend to zero volume, the quantum description shows that a bounce occurs, connecting a contracting epoch of the Universe with an expanding one.

Noticeably, the qualitative behavior coincides with that of the free system. Recall that in that case, the~expectation value of the volume is given by \eqref{eq:expVsLQC}, and can be written as a hyperbolic cosine, as has been studied in \cite{ACS}. Therefore, its trajectory describes a symmetric bounce around a point $\phi=\phi_B^F$. In the next section we investigate  the impact of the potential on the dynamics of the system further.

\section{The Effect of the Constant Potential in the Bounce Scenario of sLQC}\label{sec:bounce with potential}

In this section, we clarify the effect of the potential in the dynamics. We compare the trajectory of the volume when the cosmological constant is positive with that of the free case (without cosmological constant). More explicitly, we look for an approximate value of the point $\phi_B$ around which the bounce occurs, and seek an approximate expression for the trajectory around this point.

For simplicity, let us call $v(\phi) = \langle \hat{v} \rangle$. To find $\phi_B$, one would need to solve $v'(\phi) = 0$. For the free case, it is easily found as
\begin{equation}
    \phi_B^F = \frac{1}{2\sqrt{12\pi G}}\ln \left(\frac{v_-}{v_+}\right) + \phi_0.
\end{equation}

However, with non vanishing potential, $v'(\phi) = 0$ is a transcendental equation which cannot be solved exactly. To this end, we employ Newton Raphson's method, where, given an initial estimate to $\phi_B$, corrections to it can be found iteratively. This method is usually employed in numerical calculations to find quite accurate roots for transcendental equations. However, we use only one iteration of it analytically, in~order to find $\phi_B$ as a function of the parameters of the system, without the need for numerical substitutions. Indeed, since we are already obtaining $v(\phi)$ from a perturbative treatment in $W$, further corrections to $\phi_B$ would be irrelevant within our truncation scheme.
The initial estimate for $\phi_B$ is given by $\phi_B^F$. The first iteration of Newton Raphson's method reads $\phi_1 = \phi_B^F - v'(\phi_B^F)/v''(\phi_B^F)$. A careful calculation yields
\begin{equation}
    \phi_1 = \phi_B^F+\tilde W\left[4(\phi_B^F-\phi_0)+\frac{f(v_n)}{4\sqrt{3\pi G}}\right] + \mathcal{O}\left(\tilde{W}^2\right),
\end{equation}
where
\begin{equation}
    f(v_n) = \frac{v_{+3}+3v_-}{v_+}- \frac{v_{-3}+3v_+}{v_-} +3\left( \frac{v_{-3}v_+}{v_-^2} -\frac{v_{+3}v_-}{v_+^2} \right).
\end{equation}

Note that $\phi_1$ already includes all the corrections to first order in the potential. Indeed, the next iteration of the method would provide us with a further improved value for $\phi_B$: $\phi_2 = \phi_1 - v'(\phi_1)/v''(\phi_1)$. However we found that $v'(\phi_1)$ is of $\mathcal{O}\left(\tilde{W}^2\right)$, and so $\phi_B\simeq \phi_1$ to first order in the potential.

 It is convenient to write the evolution of the expectation value of the volume in terms of $\tilde\Delta\phi\equiv\phi-\phi_B$ instead of $\Delta\phi=\phi-\phi_0$. Expanding \eqref{eq:expv_W} in powers of $W$ and truncating at first order we find
\begin{equation}\label{eq:Vapprox}
\begin{split}
    \langle \hat V \rangle&\simeq V_B^F\bigg\{ \left[1+3\tilde W-\frac{\tilde W}{2}\left(\frac{v_{-3}+3v_+}{v_-}+\frac{v_{+3}+3v_-}{v_+} \right)\right]\cosh(\sqrt{12\pi G}\tilde\Delta\phi)\\
    &+\tilde W\left[\frac{3}{2}\left(\frac{v_{-3}v_+}{v_-^2}-\frac{v_{+3}v_-}{v_+^2}\right)-8\sqrt{3\pi G}\tilde\Delta\phi\right] \sinh(\sqrt{12\pi G}\tilde\Delta\phi)\\
    &+\frac{\tilde W}{2}\bigg[\frac{v_{+3}v_-}{v_+^2}e^{3\sqrt{12\pi G}\tilde\Delta\phi}+\frac{v_{-3}v_+}{v_-^2}e^{-3\sqrt{12\pi G}\tilde\Delta\phi}\bigg]\bigg\},
    \end{split}
\end{equation}
where $V_B^F = 4\pi G\gamma \sqrt{\Delta v_+ v_-}$ is the value of the volume at the bounce for the free system. Let us remark again that the above expression is valid for the range of the evolution when the energy density of the matter field dominates over $W=\Lambda/(8\pi G)$, and in particular around the bounce, which we describe now.

The value of the volume at the bounce for the current (non-free) system, $V_B=V(\phi_B)$, is thus given by
\begin{equation}
    V_B \simeq V_B^F \left[1+\frac{\tilde{W}}{2}\left(6-\frac{v_{+3}+3v_-}{v_+}-\frac{v_{-3}+3v_+}{v_-}+\frac{v_{+3}v_-}{v_+^2}+\frac{v_{-3}v_+}{v_-^2}\right)\right],
\end{equation}
at first order in $W$.
Equation \eqref{eq:Vapprox} allows us to analyze in more detail the effect of the potential on the dynamics. We can easily see that there is an impact on the symmetry of the bounce around $\phi_B$. Namely, we identify three different types of contributions: a symmetric one in $\tilde\Delta\phi$ from the terms multiplying $\cosh(\sqrt{12\pi G}\tilde\Delta\phi)$ and from $-8\sqrt{3\pi G}\tilde\Delta\phi\sinh(\sqrt{12\pi G}\tilde\Delta\phi)$, an anti-symmetric one given by the remaining terms multiplying $\sinh(\sqrt{12\pi G}\tilde\Delta\phi)$, and finally one that generically is neither symmetric nor anti-symmetric, given by the terms on the last line of \eqref{eq:Vapprox}, as the weights of the two exponentials do not even have to be related for generic physical states.

Whereas for generic states, the effect of the potential is somewhat intricate, remarkably, for the specific case of the semiclassical states determined by the Gaussian profiles \eqref{eq:gaussian profile}, we find that
\begin{equation}
    \frac{v_{-3}v_+}{v_-^2}=\frac{v_{+3}v_-}{v_+^2}=e^{2/\sigma^2}.
\end{equation}

Consequently, the anti-symmetric term of \eqref{eq:Vapprox} is canceled, and in the last term the coefficients multiplying the two exponentials are equal, thereby turning this into a symmetric contribution. Therefore, for this family of physical states the bounce remains symmetric, as in the case of the free system.

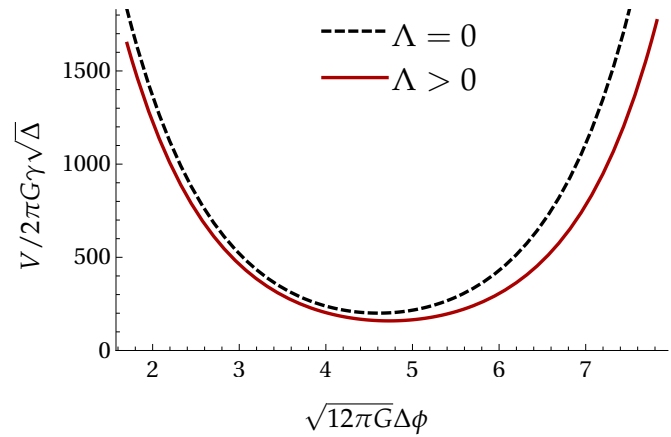
\begin{figure}[H]
\centering
  \begin{tikzpicture}
  \tikzstyle{every node}=[font=\normalsize]
  \node (img) at (0,0) {\includegraphics[width=0.5\textwidth]{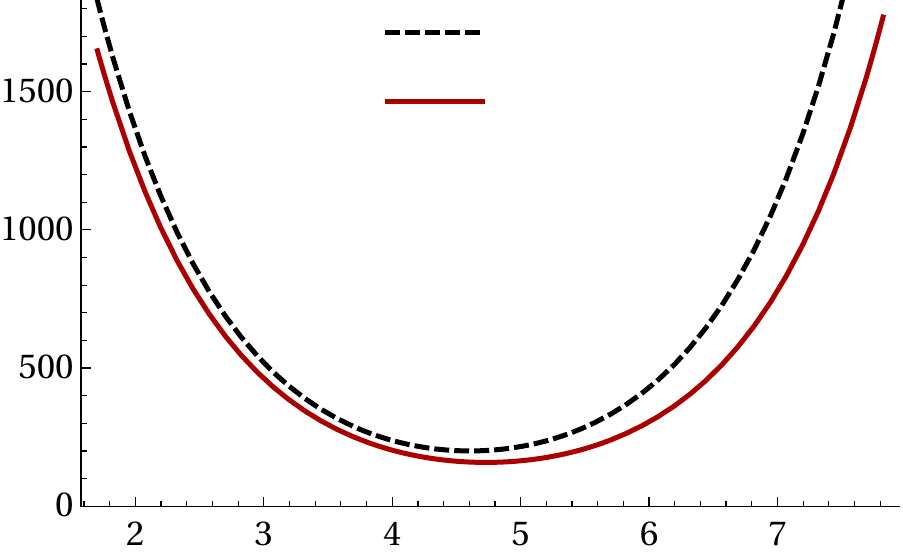}};
  \node[below=of img, node distance=0cm, yshift=1cm,] {$\sqrt{12\pi G}\Delta\phi$};
  \node[left=of img, node distance=0cm, anchor=center,rotate=90,yshift=-0.7cm,] {$V/2\pi G\gamma\sqrt{\Delta}$};
  \node[left] at (1.6,2.15) {\large{$\Lambda=0$}};
  \node[left] at (1.6,1.55) {\large{$\Lambda>0$}};
 \end{tikzpicture}
 \caption{Comparison between the expectation value of $\hat{v}$ on a semiclassical state with the profile \eqref{eq:gaussian profile}, for~the free case ($\tilde{W}=0$), and $\tilde{W} = 10^{-5}$. The parameters of the Gaussian profile were set to $k_o=-100$, $\sigma=10$ ($\langle\Delta\hat{\pi}_{\phi}\rangle/\langle\hat{\pi}_{\phi}\rangle \simeq 0.07$).}
  \label{fig:compare bounces}
\end{figure}

Figure \ref{fig:compare bounces} compares the expectation value of the volume in the two cases. For this analysis, we chose a toy value for $\tilde{W}$ so that its effect in the dynamics is visible. Indeed we see that the dynamics of the full system describes a symmetric bounce, which is displaced with respect to that of the free case.

Namely, for our class of states defined by the Gaussian profile \eqref{eq:gaussian profile}, we find that $\phi_B^F-\phi_0=\ln|k_o|/\sqrt{12\pi G}$ and $f(v_n) = \left(3-e^{2/\sigma^2}\right)(k_o^2-k_o^{-2})$. For $k_o \ll -1$ and $\sigma^{-2} \ll 1$, we obtain $e^{2/\sigma^2} \simeq 1$, and $f(v_n)\simeq 2 k_o^2$, leading to
\begin{equation}\label{eq:phiB}
    \phi_B \simeq \phi_B^F+ \frac{\tilde W}{\sqrt{12\pi G}}\left(4\ln|k_o|+k_o^2\right),
\end{equation}
and
\begin{equation}\label{eq:VB}
    V_B \simeq V_B^F \left(1-2\tilde{W} k_o^2\right).
\end{equation}
Here, $V_B^F=4\pi G\gamma\sqrt{\Delta} |k_o|e^{1/(4\sigma^2)}\simeq 4\pi G\gamma\sqrt{\Delta} |k_o|$.
Thus, we conclude that for our class of states, the effect of the potential is to push the bounce to higher $\phi$ and smaller volumes.

In summary, whereas for the specific case of semiclassical states defined with a Gaussian spectral profile, the introduction of a positive cosmological constant simply displaces the bounce by state-dependent amounts, for generic states it also renders it asymmetrical, in a state-dependent way. As a consequence, the two branches of the Universe (pre and post-bounce) tend, in general, to different classical trajectories. Even though this is the first time that an asymmetric bounce has resulted from this quantization procedure, we note that other prescriptions within the context of LQC have been found to generate an asymmetry between the pre and post-bounce epochs \cite{Assanioussi:2018,Liegener:2019,Liegener:2019_2,Li:2018}.

Furthermore, the change in the minimum of the volume also affects the maximum value of the total energy density by a state-dependent amount. However, it is not clear if the universal upper bound of the energy density of the free model found in \cite{ACS} is affected as well. Since classically, this quantity is defined as $\rho = \pi_{\phi}^2/(2 V^2)+\Lambda/(8\pi G)$, and within LQC it can also be taken as $\rho=3\sin^2(b)/(8\pi G\gamma^2\Delta)$, the definition of the quantum operator representing it is subject to factor ordering ambiguities. Thus, even though this is an interesting investigation, it requires a careful and detailed study in line with the work of~\cite{Kaminski:2008td}, which we leave for the future.

\section{Regime of Validity of Our Approximation}
\label{sec:approx}

Finally, it should be noted that the perturbative nature of this treatment implies a regime of validity with respect to the parameters of the model. Naturally, the corrections arising from the perturbative treatment should not achieve the order of the leading terms. Although one might at first assume that as long as $\tilde{W} \ll 1$ the perturbation scheme is valid, we can see from \eqref{eq:expv_W} that the corrections arise with the state-dependent variables $v_n$. Thus, there has to be some constraint on the parameters of the chosen state.
In the case of a Gaussian profile such as \eqref{eq:gaussian profile}, making use of \eqref{eq:vn_ko}, the leading contributions to the corrections are found to be of the order of $\tilde{W}k_o^2$ and $\tilde{W}k_o^2 e^{2/\sigma^2}$ with respect to the terms of the free model. Consequently, we need to restrict to the region of parameter space for which $\tilde{W}k_o^2 \ll 1$ and $\tilde{W}k_o^2 e^{2/\sigma^2} \ll 1$.

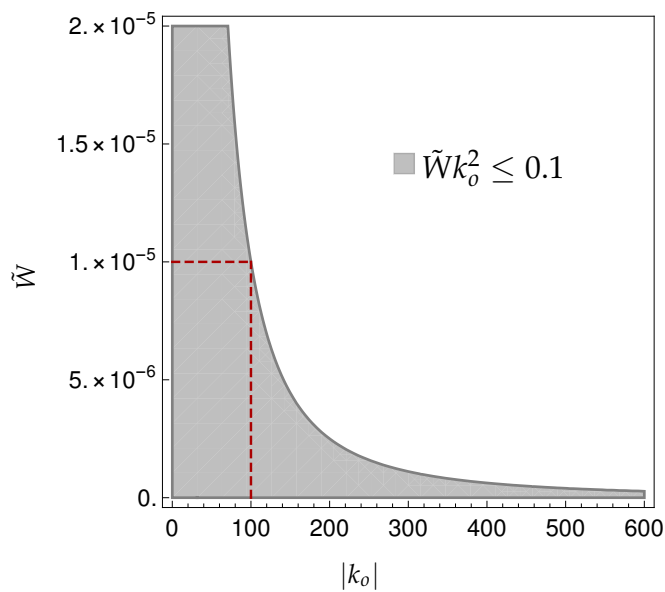
\begin{figure}[H]
\centering
  \begin{tikzpicture}
  \tikzstyle{every node}=[font=\normalsize]
  \node (img) at (0,0) {\includegraphics[width=0.5\textwidth]{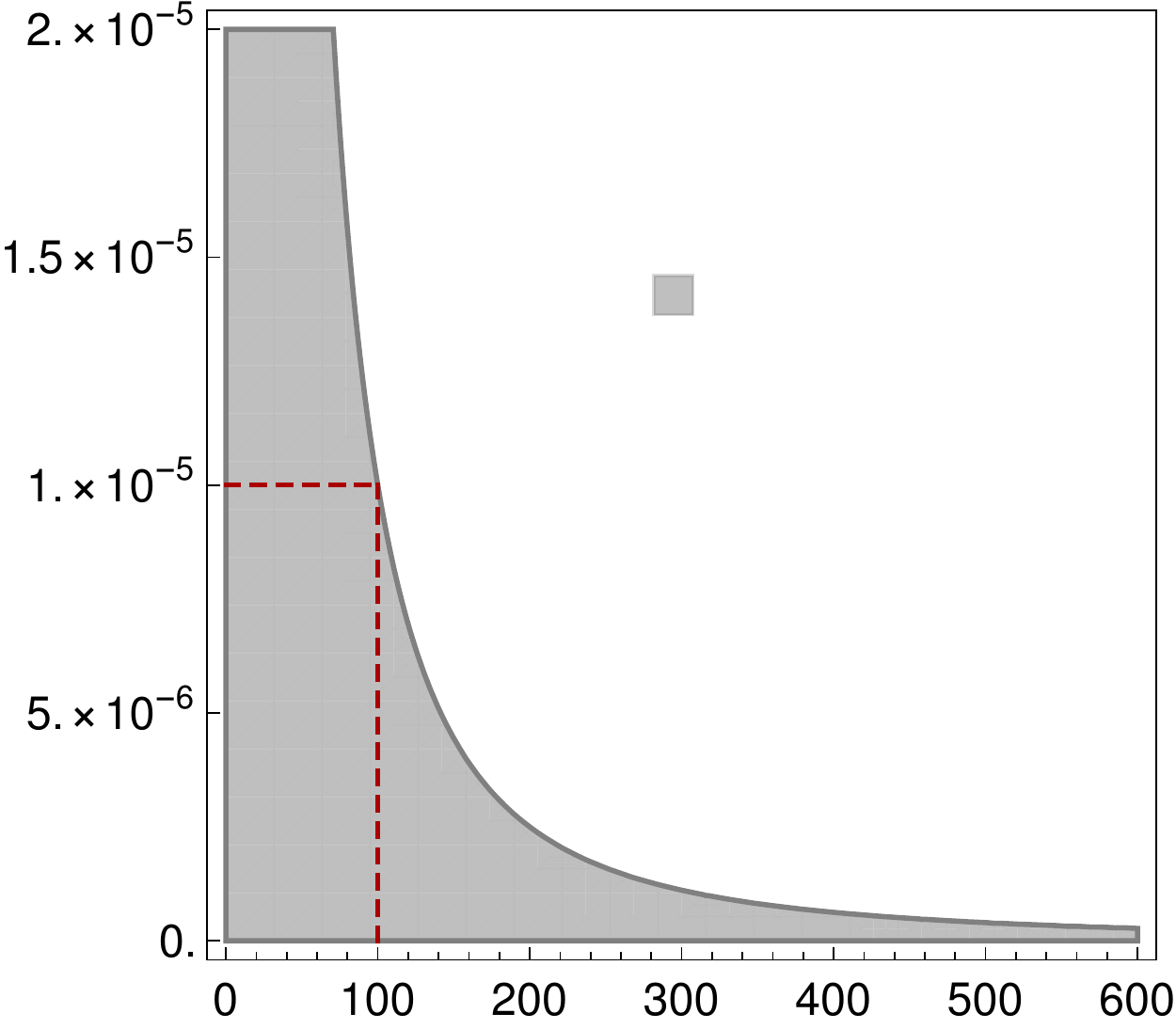}};
  \node[below=of img, node distance=0cm, yshift=1cm,] {$|k_o|$};
  \node[left=of img, node distance=0cm, anchor=center,rotate=90,yshift=-0.7cm,] {$\tilde{W}$};
  \node[left] at (2.85,1.45) {\large{$\tilde{W}k_o^2 \leq 0.1$}};;
 \end{tikzpicture}
 \caption{Representation of the region of parameters that provides a truncation error $\lesssim 1\%$, when $\sigma = 10$. The red dashed lines represent the point chosen for the graphical representations of the dynamics in this work: $|k_o| = 100$, $\tilde{W} = 10^{-5}$.}
  \label{fig:region of params}
\end{figure}

The choice of values for these three parameters requires two types of compromise. Firstly, noting that $e^{2/\sigma^2} \geq 1$, we wish to use values for $\sigma$ such that $\sigma^{-2} \ll 1$ (thus $e^{2/\sigma^2} \simeq 1$). However, at the same time, we want to keep the relative dispersion of the states given by \eqref{eq:pi_phi and disp} small, which prevents us from picking high values for $\sigma$, as these would have to be compensated for with $k_o$. This way, we constrain $\sigma$ to an intermediate value $\mathcal{O}(10)$, which allows us in what follows to choose reasonable values for the remaining parameters.
Then, with this choice for $\sigma$, the treatment is valid as long as $\tilde{W}k_o^2 \ll 1$. This is also in agreement with the condition determined in \cite{Merce_MSeq} for the validity of the treatment to first order in the potential. Consequently, the error associated to truncating the dynamics at first order in the potential is of order of $(\tilde{W} k_o^2)^2$. The values of the parameters used in the previous graphics have been chosen by limiting this error to $1\%$ at the bounce; i.e., by choosing $\tilde{W}$ and $k_o$ such that $\tilde{W} k_o^2 \leq 0.1$. Figure \ref{fig:region of params} shows the corresponding region in parameter space. In it, we highlight the point $(100,10^{-5})$ chosen for the graphical representations presented in this work. This choice was based on the second compromise: on one hand, a~larger value of $\tilde{W}$ makes more obvious the effect of the potential in the dynamics, but on the other hand, $|k_o|$ should be large enough so that the physical state has a small relative dispersion.

One might also wonder how far from the bounce we can trust our quantum evolution truncated at first order in $W$. The agreement of the quantum trajectory with the classical ones displayed in Figure~\ref{fig:dynamics} shows that the approximation remains valid well within the regime in which quantum corrections become negligible. Indeed, from the condition for validity obtained in \cite{Merce_MSeq}, we are able to determine that the approximation is valid around the bounce for $|\sqrt{12\pi G}(\Delta\phi-\phi_B^F)| \leq 1/|\tilde{W}k_o|$, which for our chosen parameters corresponds to $|\sqrt{12\pi G}(\Delta\phi-\phi_B^F)| \leq 10^3$.

\section{Conclusions/Discussion}\label{sec:discussion}

This work aimed to deepen the analysis of loop quantum gravitational corrections to pre-inflationary dynamics of cosmological models for inflation. In doing so, we intended to contribute to a falsifiable picture of LQC, as these corrections likely influence the evolution of primordial perturbations, offering a possibility of contrasting predictions of quantum cosmology (and in particular LQC) with observations from the CMB. To that end, our strategy was to develop the necessary mechanisms to employ the treatment proposed in \cite{Merce_MSeq} to flat FLRW models with a scalar field subject to a potential that drives inflation. This treatment is based on the dynamics of the model with vanishing potential, obtaining the effect of the potential in the dynamics in a perturbative manner.

In this work, we applied said method to the simplest possible case: a flat FLRW model with a scalar field subject to a constant potential. However, we showed explicit calculations with generic potential as far as was possible, so as to simplify future applications to other models.
For this specific model, equivalent to the model with a massless scalar field in the presence of a cosmological constant, we computed the expectation value of the volume along the evolution.
By plotting this quantity along with the classical trajectories, we observed the same qualitative behavior as that of the free case: far from the Planckian regime, where quantum effects are expected to be negligible, the quantum trajectory agrees with the classical ones, whereas close to the Planck scale, the quantum dynamics display a bounce, connecting a contracting epoch of the Universe with an expanding one. 

However, the main novelty of this work is that the dynamics were obtained with respect to that of the free case, making it easy to compare the two scenarios. This comparison revealed that even though the qualitative behavior of the two models (with and without cosmological constant) is the same, the bounce suffers a displacement when a cosmological constant is introduced.
Through a more detailed analysis, we~were able to write an approximate form of the expectation value of the volume around the new bounce. From it, we easily concluded that, remarkably, this simple behavior of a symmetric bounce is particular to semiclassical states defined by a Gaussian profile whose peak is essentially the expectation value of the momentum of the scalar field, which is a constant of motion. Indeed, the behavior of the volume for generic states is more intricate, as there are corrections that are not symmetric around the bounce and even ones that are anti-symmetric.
Furthermore, focusing on semiclasscial states given by Gaussian profiles, we~were able to determine that a positive cosmological constant displaces the bounce to lower values of the volume and higher values of the scalar field, by amounts that are dependent on the state parameters.

One consequence of our results, whose detailed study we leave for the future, is that the upper bound on the total energy density might be affected by a state-dependent amount. Indeed the maximum of the energy density would surely be affected in such a manner; however, it is not trivial to understand whether the universal upper bound found in the model without cosmological constant \cite{ACS} is altered as well. A~proper analysis of this point requires a study on the different possible factor orderings one may choose for the operator representing the energy density, in accordance with the discussion of \cite{Kaminski:2008td} on the spectrum of the energy density operator.

Finally, it is worth remarking that even though this treatment introduces approximations, it offers several advantages. Firstly, it is less time and resource-consuming than the exact treatment pursued in~\cite{Pawlowski2012}. Secondly, it can be applied to models with non-constant potentials, presumably with little in the way of added difficulties, other than the computation of one integral that depends on the form of the field potential. To this, we add that at least polynomial potentials in the scalar field should not introduce additional challenges. Lastly, the application of this method to a simple model already allowed us to uncover interesting consequences that went unnoticed in the numerical investigations of \cite{Pawlowski2012}. Indeed, even though the treatment of that work was more exact, the form of the physical states was so intricate that it did not provide information on the dynamics for generic states. A particularization to semiclassical states is necessary and some state-dependent effects are naturally ignored.

In the future, we intend to apply this treatment to other more interesting forms of the potential that drive inflation, obtaining further quantum corrections to their dynamics. Then, primordial perturbations can be imposed on these corrected backgrounds, and predictions of power-spectra can be computed, so~that a contrast of predictions from quantum cosmology with observations from the CMB is possible.

\vspace{6pt} 




\authorcontributions{The authors have contributed equally to this work and have read and agreed to the published version of the manuscript.} 

\funding{Financial support was provided by the Spanish Government through the project  FIS2017-86497-C2-2-P (with FEDER contribution).
R. Neves acknowledges financial support from Funda\c{c}\~ao para a Ci\^encia e a Tecnologia (FCT) through the research grant SFRH/BD/143525/2019.}

\acknowledgments{We thank J. Olmedo for useful discussions.}

\conflictsofinterest{The authors declare no conflict of interest.} 

%
%
\appendixtitles{yes} 
\appendix
\appendixsections{multiple}
\section{Calculations of the Volume for Constant Potential}\label{sec:appendix_V}

In section \ref{sec:expV_constantW}, we have laid out the necessary steps to compute $\langle\hat{V}\rangle$ for a given form of the potential $W(\phi)$. Let us do so for a constant potential $W(\phi)=W$. With this potential, one obtains $I_1$ defined in~\eqref{eq:I1}~as:
\begin{equation}
    I_1(x,\phi) = \frac{1}{2}W\left\{\Delta\phi+\frac{1}{4\sqrt{3\pi G}}\bigg[\sinh\left(4\sqrt{3\pi G}x\right)-\sinh\left(4\sqrt{3\pi G}(x-\Delta\phi)\right)\bigg]\right\},
\end{equation}
which leads to
\begin{equation}
    \partial_xI_1(x,\phi) = \frac{1}{2}W\bigg[\cosh\left(4\sqrt{3\pi G}x\right)-\cosh\left(4\sqrt{3\pi G}(x-\Delta\phi)\right)\bigg].
\end{equation}

Consequently, we obtain $E(x,\phi)$ as defined in \eqref{eq:E}:
\begin{equation}
    \begin{split}
        E(x,\phi)&=\left(1+3\tilde{W}\right)\cosh\left(\sqrt{12\pi G}(x-\Delta\phi)\right)+\tilde{W}\Bigg[8\sqrt{3\pi G}\Delta\phi \sinh\left(\sqrt{12\pi G}(x-\Delta\phi)\right)\\
        &+\cosh\left(3\sqrt{12\pi G}(x-\Delta\phi)\right) -\cosh\left(\sqrt{12\pi G}(3x-\Delta\phi)\right) -3\cosh\left(\sqrt{12\pi G}(x+\Delta\phi)\right)\Bigg],
    \end{split}
\end{equation}
where, we recall, $\tilde{W}$ is defined in \eqref{eq:Wtilde}. 

Once $E(x,\phi)$ is inserted in \eqref{eq:VexpW}, the terms of the form
\begin{equation}
    \cosh\left[\sqrt{12\pi G}\left(n_x x-n_{\phi}\phi\right)\right] 
\end{equation}
lead to contributions of the form
\begin{align}
  \frac{1}{2} \bigg[v_{-n_x}e^{-n_{\phi}\sqrt{12\pi G} \phi}+v_{+n_x}e^{n_{\phi}\sqrt{12\pi G} \phi}\bigg],
\end{align}
and similarly for the $\sinh$ terms (changing the sign of the second term in the above expression).

Finally, this leads to $\langle \hat{V} \rangle$, as defined in \eqref{eq:expv_W}.

With the same strategy, we also obtain $\langle\hat{V}^2\rangle$. Namely, we find $\hat{V}^2_J$ from \eqref{eq:Aj}, by using $\hat V_I^2$ in place of $\hat{A}_I$. Then, as in \eqref{eq:Vexp1}, we calculate
\begin{align}
    \langle \hat{V}^2 \rangle\simeq \langle \hat{V}^2_J \rangle =2i\int_{-\infty}^\infty dx \frac{dF^*(x)}{dx}\hat{V}^2_J(\phi)F(x),
\end{align}
arriving to the result
\begin{align}
\begin{split}
    \langle\hat{V}^2\rangle&\simeq\left(2\pi G\gamma\sqrt{\Delta}\right)^2\Bigg[ \left(\frac{v_{d_{+2}}^2}{2} +\tilde{W} w_{2_+}(\phi)\right) e^{2\sqrt{12\pi G}\Delta\phi}+ \left(\frac{v_{d_{-2}}^2}{2} +\tilde{W} w_{2_-}(\phi)\right) e^{-2\sqrt{12\pi G}\Delta\phi}\\
    & + \tilde{W}\left(v_{d_{+4}}^2e^{4\sqrt{12\pi G}\Delta\phi}+ v_{d_{-4}}^2e^{-4\sqrt{12\pi G}\Delta\phi}\right)+\left(1+6\tilde{W}\right)v_{d_0}^2-4\tilde{W}\left(v_{d_{+2}}^2+v_{d_{-2}}^2\right)\Bigg],
\end{split}
\end{align}
where
\begin{align}
    w_{2_\pm}(\phi) &\equiv \left(4\mp 8\sqrt{3\pi G}\Delta\phi\right)v_{d_{\pm2}}^2 - v_{d_{\pm4}}^2-3v_{d_0}^2,\\
    v_{d_n} &\equiv \frac{1}{3\pi G}\int_{-\infty}^{+\infty} dx \text{Im}\left[ \frac{dF^*(x)}{dx}\frac{d^2F(x)}{dx^2} \right] e^{-n\sqrt{12\pi G}\ x}.
\end{align}

This allows the computation of the relative dispersion of $\hat{V}$, ${\langle\Delta\hat{V}\rangle}/{\langle\hat{V}\rangle} = \sqrt{{\langle\hat{V}^2\rangle}/{\langle\hat{V}\rangle^2}-1}$. We~truncate to first order in $W$, getting a long expression whose form  we omit for simplicity.
\externalbibliography{yes}
\bibliography{LQCbib}

\begin{thebibliography}{-------}
\providecommand{\natexlab}[1]{#1}

\bibitem[Ade and \emph{et al}(2016{\natexlab{a}})]{Ade_2015PlanckResults}
Ade, P.A.R.; \emph{et al}.
\newblock {Planck 2015 results. XIII. Cosmological parameters}.
\newblock {\em Astron. Astrophys.} {\bf 2016}, {\em 594},~A13.

\bibitem[Ade and \emph{et al}(2016{\natexlab{b}})]{Ade_2015PlanckResults2}
Ade, P.A.R.; \emph{et al}.
\newblock {Planck 2015 results. XX. Constraints on inflation}.
\newblock {\em Astron. Astrophys.} {\bf 2016}, {\em 594},~A20.

\bibitem[Kiefer and Kr\"amer(2012)]{Kiefer:2012}
Kiefer, C.; Kr\"amer, M.
\newblock Quantum Gravitational Contributions to the Cosmic Microwave
  Background Anisotropy Spectrum.
\newblock {\em Phys. Rev. Lett.} {\bf 2012}, {\em 108},~021301.

\bibitem[Brizuela \em{et~al.}(2016)Brizuela, Kiefer, and
  Kr\"amer]{Brizuela:2016}
Brizuela, D.; Kiefer, C.; Kr\"amer, M.
\newblock Quantum-gravitational effects on gauge-invariant scalar and tensor
  perturbations during inflation: The de Sitter case.
\newblock {\em Phys. Rev. D} {\bf 2016}, {\em 93},~104035.
\newblock
  doi:{\changeurlcolor{black}\href{https://doi.org/10.1103/PhysRevD.93.104035}{\detokenize{10.1103/PhysRevD.93.104035}}}.

\bibitem[Bojowald(2005)]{LQCreview_Bojowald2005}
Bojowald, M.
\newblock {Loop quantum cosmology}.
\newblock {\em Living Rev. Rel.} {\bf 2005}, {\em 8},~11.

\bibitem[Ashtekar and Singh(2011)]{LQCreview_Ashtekar2011}
Ashtekar, A.; Singh, P.
\newblock {Loop Quantum Cosmology: A Status Report}.
\newblock {\em Class. Quant. Grav.} {\bf 2011}, {\em 28},~213001.

\bibitem[Banerjee \em{et~al.}(2012)Banerjee, Calcagni, and
  Martin-Benito]{LQCreview_Banerjee2012}
Banerjee, K.; Calcagni, G.; Martin-Benito, M.
\newblock {Introduction to loop quantum cosmology}.
\newblock {\em SIGMA} {\bf 2012}, {\em 8},~016.

\bibitem[Agullo and Singh(2017)]{LQCreview_Agullo2016}
Agullo, I.; Singh, P.
\newblock {Loop Quantum Cosmology}. In {\em Loop Quantum Gravity: The First 30
  Years}; Ashtekar, A.; Pullin, J., Eds.; WSP,  2017; pp. 183--240.

\bibitem[Ashtekar \em{et~al.}(2006{\natexlab{a}})Ashtekar, Pawlowski, and
  Singh]{APS_PRL}
Ashtekar, A.; Pawlowski, T.; Singh, P.
\newblock {Quantum nature of the big bang}.
\newblock {\em Phys. Rev. Lett.} {\bf 2006}, {\em 96},~141301.

\bibitem[Ashtekar \em{et~al.}(2006{\natexlab{b}})Ashtekar, Pawlowski, and
  Singh]{APS_extended}
Ashtekar, A.; Pawlowski, T.; Singh, P.
\newblock {Quantum Nature of the Big Bang: Improved dynamics}.
\newblock {\em Phys. Rev.} {\bf 2006}, {\em D74},~084003.

\bibitem[Bentivegna and Pawlowski(2008)]{Bentivegna2008}
Bentivegna, E.; Pawlowski, T.
\newblock {Anti-deSitter universe dynamics in LQC}.
\newblock {\em Phys. Rev.} {\bf 2008}, {\em D77},~124025.

\bibitem[Kaminski and Pawlowski(2010)]{Kaminski2009}
Kaminski, W.; Pawlowski, T.
\newblock {The LQC evolution operator of FRW universe with positive
  cosmological constant}.
\newblock {\em Phys. Rev.} {\bf 2010}, {\em D81},~024014.

\bibitem[Pawlowski and Ashtekar(2012)]{Pawlowski2012}
Pawlowski, T.; Ashtekar, A.
\newblock {Positive cosmological constant in loop quantum cosmology}.
\newblock {\em Phys. Rev.} {\bf 2012}, {\em D85},~064001.

\bibitem[Mart\'{\i}n-Benito \em{et~al.}(2008)Mart\'{\i}n-Benito, Garay, and
  Marug\'an]{Merce_2008Hybrid}
Mart\'{\i}n-Benito, M.; Garay, L.J.; Marug\'an, G.A.M.
\newblock Hybrid quantum Gowdy cosmology: Combining loop and Fock
  quantizations.
\newblock {\em Phys. Rev. D} {\bf 2008}, {\em 78},~083516.

\bibitem[Garay \em{et~al.}(2010)Garay, Mart\'{\i}n-Benito, and
  Mena~Marug\'an]{Garay_2008hybrid}
Garay, L.J.; Mart\'{\i}n-Benito, M.; Mena~Marug\'an, G.A.
\newblock Inhomogeneous loop quantum cosmology: Hybrid quantization of the
  Gowdy model.
\newblock {\em Phys. Rev. D} {\bf 2010}, {\em 82},~044048.

\bibitem[Mart\'{\i}n-Benito \em{et~al.}(2010)Mart\'{\i}n-Benito,
  Mena~Marug\'an, and Wilson-Ewing]{Merce_2010hybrid}
Mart\'{\i}n-Benito, M.; Mena~Marug\'an, G.A.; Wilson-Ewing, E.
\newblock Hybrid quantization: From Bianchi I to the Gowdy model.
\newblock {\em Phys. Rev. D} {\bf 2010}, {\em 82},~084012.

\bibitem[Fern\'andez-M\'endez \em{et~al.}(2012)Fern\'andez-M\'endez,
  Mena~Marug\'an, and Olmedo]{Mendez_2012hybridInflation}
Fern\'andez-M\'endez, M.; Mena~Marug\'an, G.A.; Olmedo, J.
\newblock Hybrid quantization of an inflationary universe.
\newblock {\em Phys. Rev. D} {\bf 2012}, {\em 86},~024003.

\bibitem[Fern\'andez-M\'endez \em{et~al.}(2014)Fern\'andez-M\'endez,
  Mena~Marug\'an, and Olmedo]{Mendez_2014hybridInflation}
Fern\'andez-M\'endez, M.; Mena~Marug\'an, G.A.; Olmedo, J.
\newblock Effective dynamics of scalar perturbations in a flat
  Friedmann-Robertson-Walker spacetime in loop quantum cosmology.
\newblock {\em Phys. Rev. D} {\bf 2014}, {\em 89},~044041.

\bibitem[Gomar \em{et~al.}(2014)Gomar, Fern\'andez-M\'endez, Marug\'an, and
  Olmedo]{Gomar_2014hybridInflation}
Gomar, L.C.; Fern\'andez-M\'endez, M.; Marug\'an, G.A.M.; Olmedo, J.
\newblock Cosmological perturbations in hybrid loop quantum cosmology:
  Mukhanov-Sasaki variables.
\newblock {\em Phys. Rev. D} {\bf 2014}, {\em 90},~064015.

\bibitem[Gomar \em{et~al.}(2015)Gomar, Martín-Benito, and
  Marugán]{Gomar_2015hybridInflation}
Gomar, L.C.; Martín-Benito, M.; Marugán, G.A.M.
\newblock {Gauge-Invariant Perturbations in Hybrid Quantum Cosmology}.
\newblock {\em JCAP} {\bf 2015}, {\em 1506},~045.

\bibitem[Agullo \em{et~al.}(2013)Agullo, Ashtekar, and Nelson]{Agullo:2013ai}
Agullo, I.; Ashtekar, A.; Nelson, W.
\newblock {The pre-inflationary dynamics of loop quantum cosmology: Confronting
  quantum gravity with observations}.
\newblock {\em Class. Quant. Grav.} {\bf 2013}, {\em 30},~085014.

\bibitem[Agullo and Morris(2015)]{Agullo_2015PS}
Agullo, I.; Morris, N.A.
\newblock Detailed analysis of the predictions of loop quantum cosmology for
  the primordial power spectra.
\newblock {\em Phys. Rev. D} {\bf 2015}, {\em 92},~124040.

\bibitem[Agullo(2015)]{Agullo_2015CMB}
Agullo, I.
\newblock Loop quantum cosmology, non-Gaussianity, and CMB power asymmetry.
\newblock {\em Phys. Rev. D} {\bf 2015}, {\em 92},~064038.

\bibitem[Ashtekar and Gupt(2017)]{Ashtekar:2016wpi}
Ashtekar, A.; Gupt, B.
\newblock {Quantum Gravity in the Sky: Interplay between fundamental theory and
  observations}.
\newblock {\em Class. Quant. Grav.} {\bf 2017}, {\em 34},~014002.

\bibitem[Martín~de Blas and Olmedo(2016)]{deBlas:2016puz}
Martín~de Blas, D.; Olmedo, J.
\newblock {Primordial power spectra for scalar perturbations in loop quantum
  cosmology}.
\newblock {\em JCAP} {\bf 2016}, {\em 1606},~029.

\bibitem[Castelló~Gomar \em{et~al.}(2017)Castelló~Gomar, Mena~Marugán,
  Martín De~Blas, and Olmedo]{Gomar:2017yww}
Castelló~Gomar, L.; Mena~Marugán, G.A.; Martín De~Blas, D.; Olmedo, J.
\newblock {Hybrid loop quantum cosmology and predictions for the cosmic
  microwave background}.
\newblock {\em Phys. Rev.} {\bf 2017}, {\em D96},~103528.

\bibitem[Castell\'o~Gomar \em{et~al.}(2016)Castell\'o~Gomar, Mart\'in-Benito,
  and Mena~Marug\'an]{Merce_MSeq}
Castell\'o~Gomar, L.; Mart\'in-Benito, M.; Mena~Marug\'an, G.A.
\newblock {Quantum corrections to the Mukhanov-Sasaki equations}.
\newblock {\em Phys. Rev.} {\bf 2016}, {\em D93},~104025.

\bibitem[Ashtekar \em{et~al.}(2008)Ashtekar, Corichi, and Singh]{ACS}
Ashtekar, A.; Corichi, A.; Singh, P.
\newblock {Robustness of key features of loop quantum cosmology}.
\newblock {\em Phys. Rev.} {\bf 2008}, {\em D77},~024046.

\bibitem[Martín-Benito and Neves(2019)]{paper1_domainV}
Martín-Benito, M.; Neves, R.B.
\newblock {Solvable Loop Quantum Cosmology: domain of the volume observable and
  semiclassical states}.
\newblock {\em Phys. Rev.} {\bf 2019}, {\em D99},~043525.

\bibitem[Assanioussi \em{et~al.}(2018)Assanioussi, Dapor, Liegener, and
  Paw\l{}owski]{Assanioussi:2018}
Assanioussi, M.; Dapor, A.; Liegener, K.; Paw\l{}owski, T.
\newblock Emergent de Sitter Epoch of the Quantum Cosmos from Loop Quantum
  Cosmology.
\newblock {\em Phys. Rev. Lett.} {\bf 2018}, {\em 121},~081303.
\newblock
  doi:{\changeurlcolor{black}\href{https://doi.org/10.1103/PhysRevLett.121.081303}{\detokenize{10.1103/PhysRevLett.121.081303}}}.

\bibitem[Liegener and Singh(2019{\natexlab{a}})]{Liegener:2019}
Liegener, K.; Singh, P.
\newblock Some physical implications of regularization ambiguities in SU(2)
  gauge-invariant loop quantum cosmology.
\newblock {\em Phys. Rev. D} {\bf 2019}, {\em 100},~124049.
\newblock
  doi:{\changeurlcolor{black}\href{https://doi.org/10.1103/PhysRevD.100.124049}{\detokenize{10.1103/PhysRevD.100.124049}}}.

\bibitem[Liegener and Singh(2019{\natexlab{b}})]{Liegener:2019_2}
Liegener, K.; Singh, P.
\newblock New loop quantum cosmology modifications from gauge-covariant fluxes.
\newblock {\em Phys. Rev. D} {\bf 2019}, {\em 100},~124048.
\newblock
  doi:{\changeurlcolor{black}\href{https://doi.org/10.1103/PhysRevD.100.124048}{\detokenize{10.1103/PhysRevD.100.124048}}}.

\bibitem[Li \em{et~al.}(2018)Li, Singh, and Wang]{Li:2018}
Li, B.F.; Singh, P.; Wang, A.
\newblock Towards cosmological dynamics from loop quantum gravity.
\newblock {\em Phys. Rev. D} {\bf 2018}, {\em 97},~084029.
\newblock
  doi:{\changeurlcolor{black}\href{https://doi.org/10.1103/PhysRevD.97.084029}{\detokenize{10.1103/PhysRevD.97.084029}}}.

\bibitem[Kaminski \em{et~al.}(2009)Kaminski, Lewandowski, and
  Pawlowski]{Kaminski:2008td}
Kaminski, W.; Lewandowski, J.; Pawlowski, T.
\newblock {Physical time and other conceptual issues of QG on the example of
  LQC}.
\newblock {\em Class. Quant. Grav.} {\bf 2009}, {\em 26},~035012.

\end{thebibliography}

\end{document}